\documentclass{article}
\usepackage{frascatiphys}
\usepackage{graphicx}
\newcommand{\beq}{\begin{equation}}
\newcommand{\eeq}{\end{equation}}
\newcommand{\bea}{\begin{eqnarray}}
\newcommand{\eea}{\end{eqnarray}}

\begin{document}
\title{ 
PERSPECTIVES OF PHOTON PHYSICS AT FUTURE COLLIDERS
}
\author{
Michael Klasen\\
{\em Institut f\"ur Theoretische Physik, Westf\"alische
 Wilhelms-Universit\"at M\"unster,} \\ {\em Wilhelm-Klemm-Stra\ss{}e 9,
 48149 M\"unster, Germany}
}
\maketitle
\baselineskip=11.6pt
\begin{abstract}
We review current results on physics with photons at the LHC and
discuss realistic perspectives of photon physics at future colliders. In
particular, we focus on Standard Model (SM) measurements with photons at the
upcoming high-luminosity and a possible high-energy LHC as well as jet
measurements at an Electron-Ion Collider (EIC) to be constructed either at BNL
or at JLAB and their potential to constrain nuclear parton densities. We also
discuss future searches for physics beyond the SM with photons in the
high-luminosity phase of the LHC.
\end{abstract}
\baselineskip=14pt
%

\section{Motivation}

Discussions of physics with photons at future colliders have traditionally
focused on photon-photon interactions at linear $e^+e^-$ colliders like TESLA
or CLIC \cite{Badelek:2001xb}, their low background, and their superior
precision for measurements of the properties of the Higgs boson or yet to be
discovered physics beyond the Standard Model (BSM) like supersymmetric (SUSY)
particles \cite{Berge:2000cb}. Unfortunately, the Large Hadron Collider (LHC)
has so far produced no evidence for BSM particles, so that the
decision to build a linear collider is still pending. The LHC has, however,
produced many interesting events with single photons, diphotons and photons
plus jets in pp, pPb and PbPb collisions. They have led to a large variety of
results ranging from the discovery of the Higgs boson \cite{Aad:2012tfa}
to the determination of the effective temperature of the quark-gluon plasma
(QGP) \cite{Klasen:2013mga}. In addition,
ultraperipheral collisions (UPCs) at the LHC have led to measurements of
exclusive dilepton and quarkonium photoproduction and even the discovery of
light-by-light scattering \cite{Aaboud:2017bwk}.

The upgrade of the LHC to its high-luminosity (HL) phase is currently underway,
and plans are being made to install stronger magnets in the existing tunnel for
a high-energy (HE) machine with increased centre-of-mass energy from 13 TeV to
as much as 27 TeV in pp, 17 TeV in pPb and 10.6 TeV in PbPb collisions. At the
same time, plans to supplement the existing Relativistic Heavy Ion Collider
(RHIC) at BNL with a circular electron accelerator or to extend the upgraded
Continuous Electron Beam Accelerator Facility (CEBAF) at JLAB with a heavy-ion
storage ring are well advanced. In both incarnations, such an Electron Ion
Collider (EIC) would greatly improve our knowledge of nuclear matter, probed
by the real and virtual photons emitted from the electron. It is therefore
appropriate to explore the impact of photons in these realistic future collider
scenarios, i.e.\ on future SM studies at the HL/HE LHC and EIC as well as
on BSM physics at the HL LHC.

\section{SM physics with photons at the high-luminosity and high-energy LHC}

\subsection{Prompt photon production}

The LHC collaborations ALICE, ATLAS and CMS have recently produced a large
variety of interesting prompt photon results in pp, pPb and PbPb collisions
at different centre-of-mass energies \cite{ns}. They serve to test
both the QCD and electroweak sectors of the SM, to constrain the parton
distribution functions (PDFs) in protons and nuclei and to determine the
background for new physics searches with ever higher precision. To fully
exploit the potential of these data, one must not only cleanly eliminate the
meson decay contributions by data-driven subtraction methods or with
infrared-safe photon isolation criteria, but also confront them with
theoretical calculations at next-to-next-to-leading order (NNLO) or using
resummation and parton showers (PS) \cite{mh}.
\begin{figure}
 \includegraphics[width=0.49\textwidth]{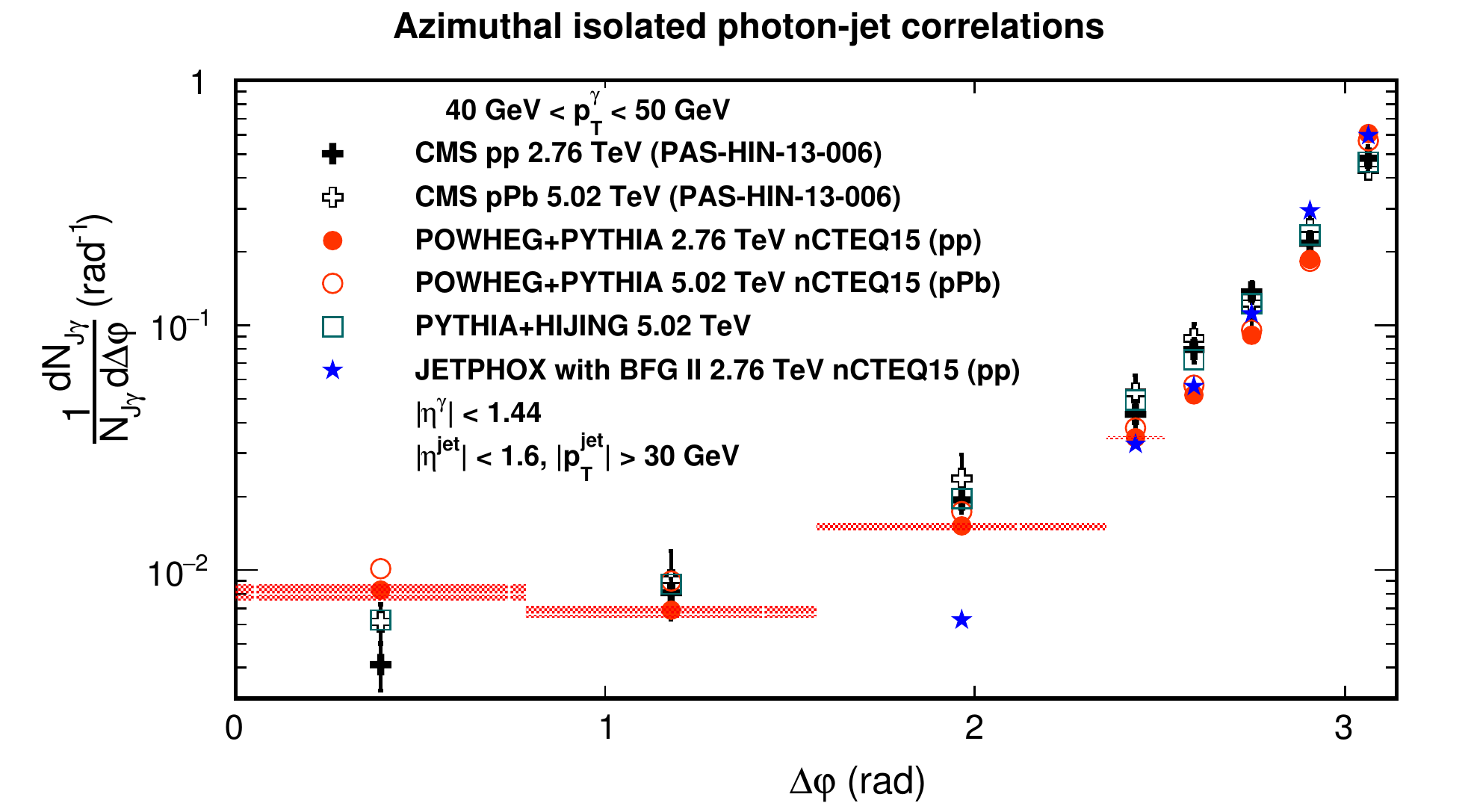}
 \includegraphics[width=0.49\textwidth]{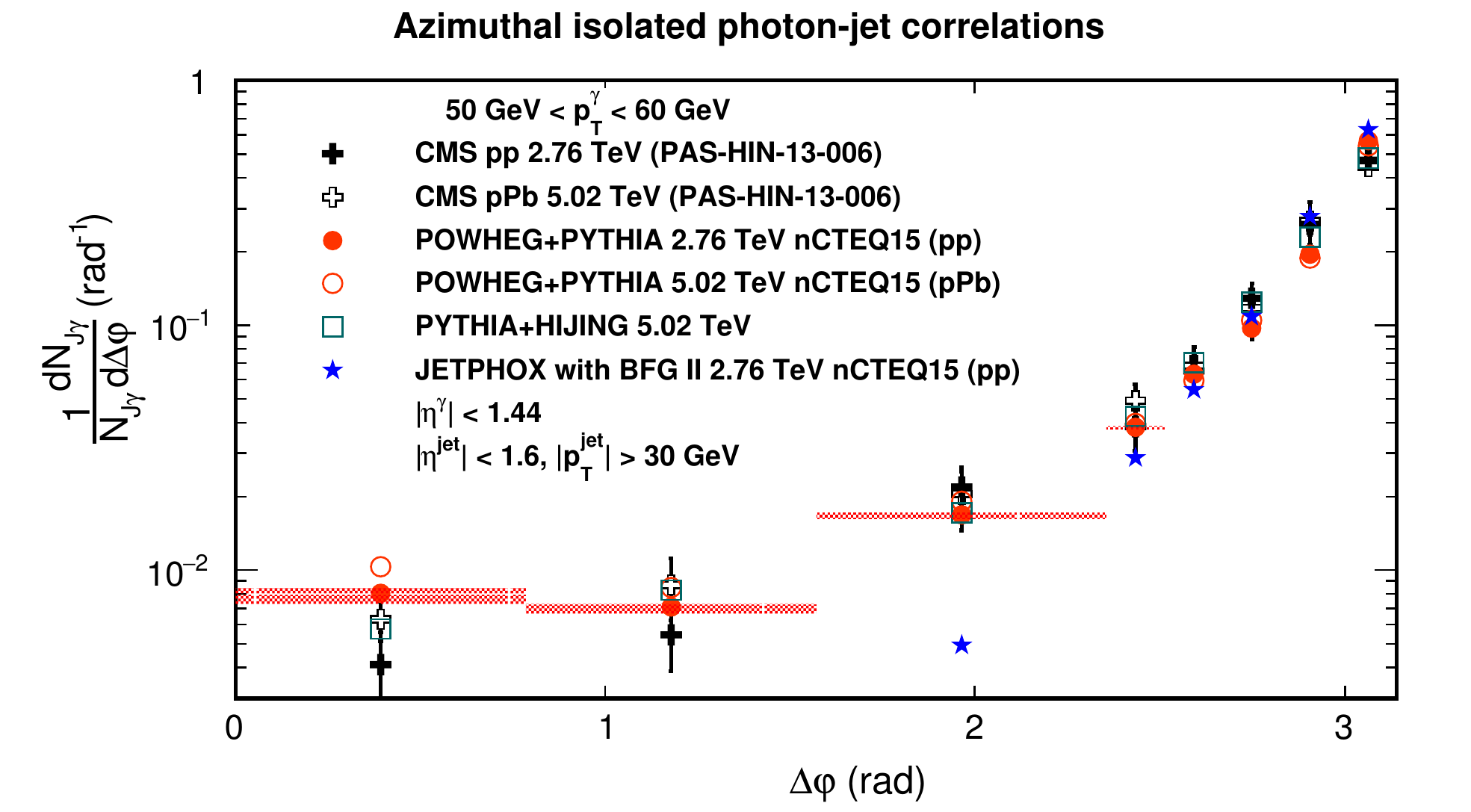}
 \includegraphics[width=0.49\textwidth]{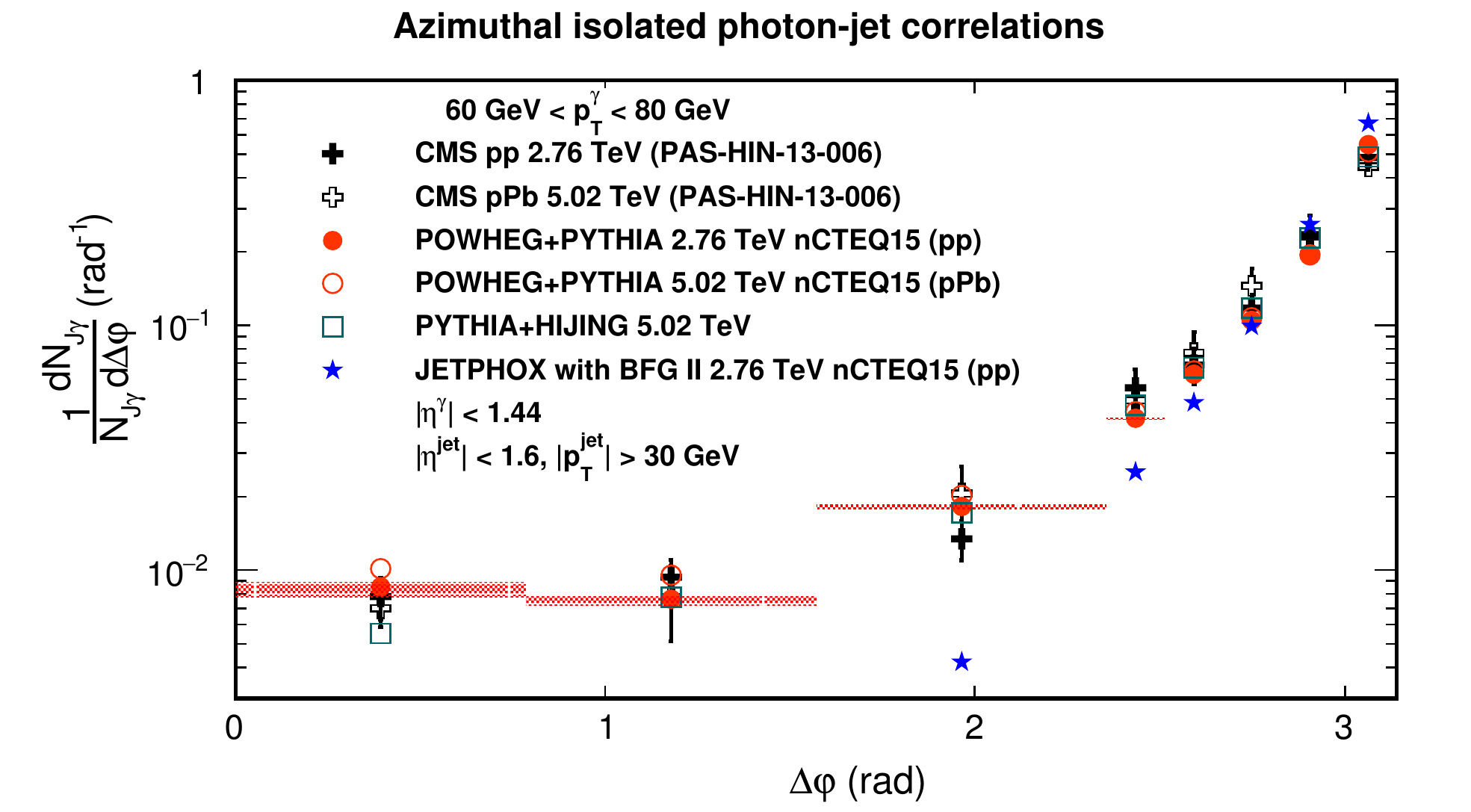}\,\,\,
 \includegraphics[width=0.49\textwidth]{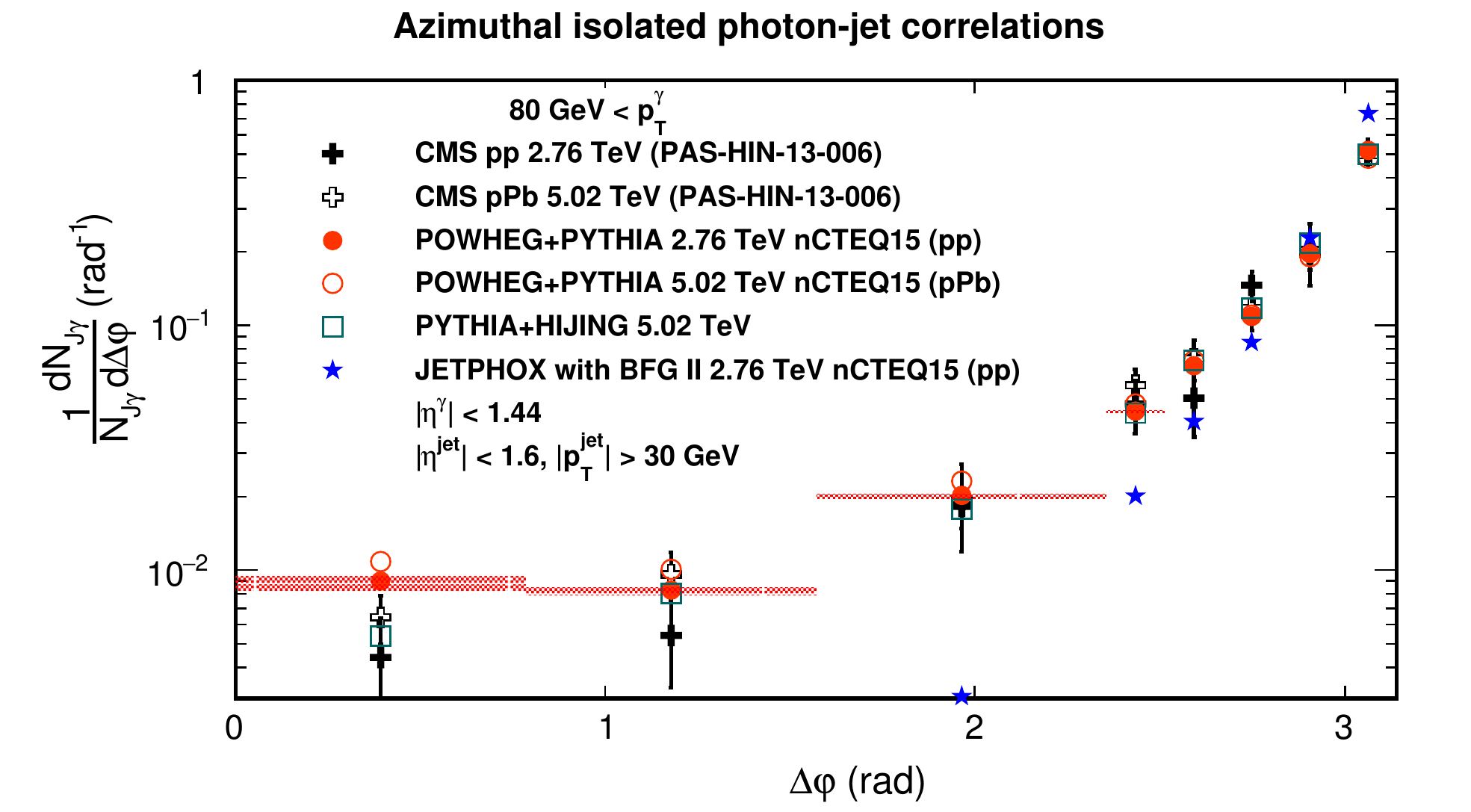}
 \caption{Relative azimuthal angle distributions of jets and
 photons in pp (open circles) and pPb (full circles) collisions at the LHC
 with a centre-of-mass energy per nucleon of
 $\sqrt{s_{NN}}=2.76$ and 5.02 TeV, respectively, in four different bins of
 photon transverse momentum. CMS 2013 data (black) \cite{CMS:2013oua} are
 compared to predictions in LO with PYTHIA+HIJING (green), NLO with JETPHOX
 (blue), and NLO+PS with POWHEG+PYTHIA (red) \cite{Jezo:2016ypn}.}
\end{figure}

This has recently been demonstrated with the implementation of photons
in POWHEG, the successful application of this new calculation to data from
ALICE, ATLAS and CMS, and predictions for future measurements with LHCb
\cite{Jezo:2016ypn}. An observable that is particularly sensitive to QCD
effects beyond next-to-leading order (NLO) is the photon-jet azimuthal
distribution, measured by CMS in pp and pPb collisions \cite{CMS:2013oua} and
shown in Fig.\ 1. While the NLO JETPHOX calculations do not describe the data
below $\Delta\phi_{J\gamma}=2\pi/3$, as the maximum number of jets is limited
to two at NLO, the POWHEG predictions agree quite well with the data. No
significant energy dependence or cold nuclear effects are yet observed with
this limited statistics, making its increase at the HL LHC mandatory.
Exploratory studies have shown that the HL LHC can reach inclusive photon and
diphoton transverse momenta up to 5 and 2 TeV, i.e.\ far beyond the current
reach of about 1 TeV and 700 GeV, respectively \cite{Azzi:2019yne}, and the
kinematic reach would obviously be even larger at a HE LHC. This would give
access to proton PDFs over a wide range in $x$ from less than $10^{-4}$ to
0.5. 

\subsection{Inclusive photoproduction}

Collisions with an tagged proton or intact nucleus,
small multiplicity, or a substantial rapidity gap on one side of the detector
system allow for the identification
of inclusive photoproduction events and thus the use of the LHC as a
photon-proton or photon-ion collider. The inclusive photoproduction of dijets
has already been observed by ATLAS \cite{Angerami} and been shown to agree
well with NLO QCD calculations \cite{Guzey:2018dlm}. With a future precision
of 5\%, these data would have the potential to reduce the nuclear PDF
uncertainties at $x\sim10^{-3}$ by more than a factor of two
\cite{Guzey:2019kik}.

\begin{figure}
 \centering
 \includegraphics[width=0.85\textwidth]{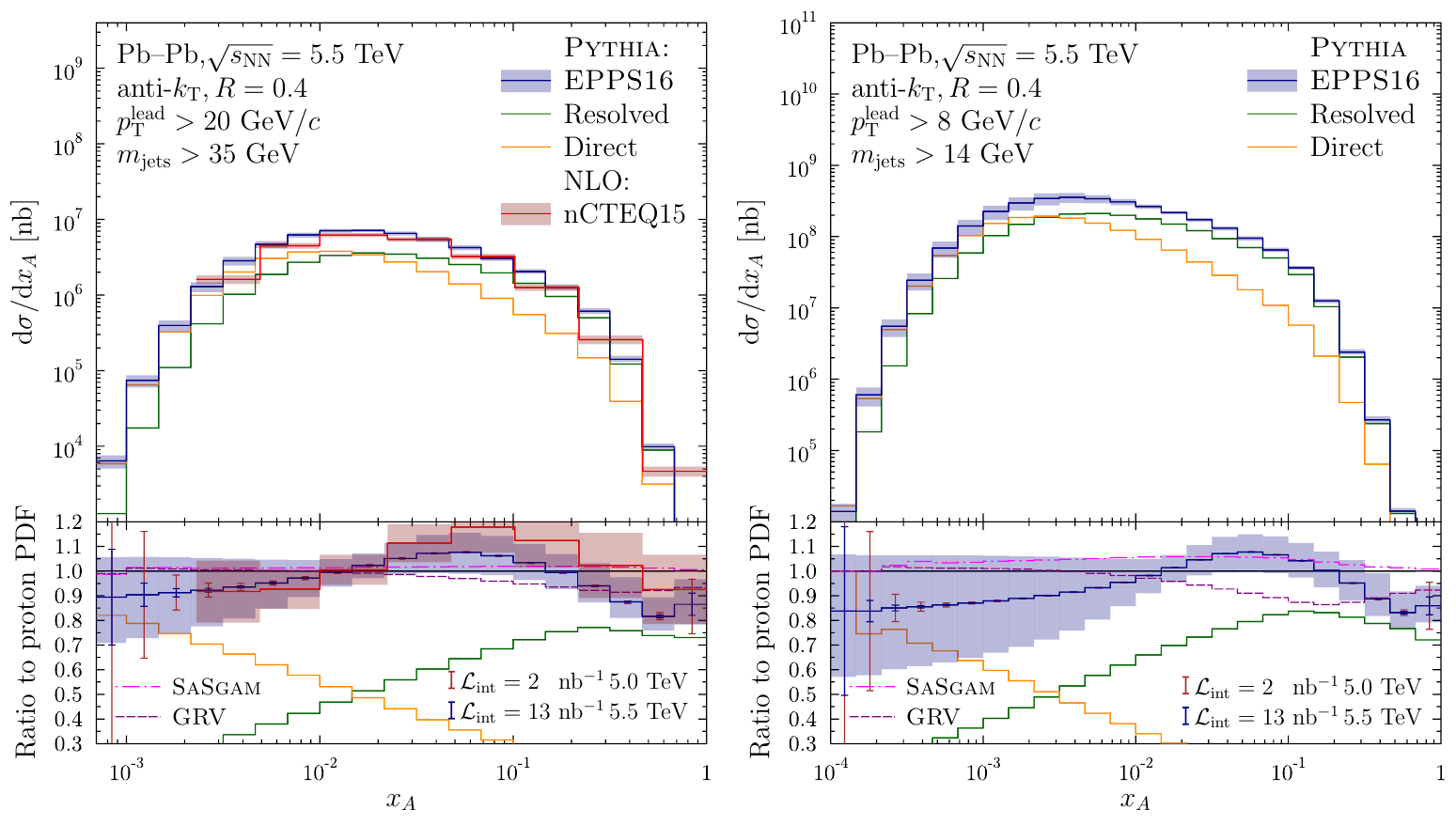}
 \caption{Photonuclear dijet cross sections in ultraperipheral PbPb
 collisions at $\sqrt{s_{NN}} = 5.5$ TeV with leading jet $p_T$ cut of 20 GeV
 (left) and 8 GeV (right). Results based on PYTHIA simulations are calculated
 with EPPS16 nuclear modification (blue), and the contributions from resolved
 (green) and direct (orange) photons are separately shown. Ratio plots show
 also results with different photon PDF sets and the expected statistical
 uncertainties corresponding to the LHC (brown) and the Run 3 and Run 4
 (dark blue) luminosities. Corresponding results based on NLO calculations
 for PbPb collisions at $\sqrt{s_{NN}} = 5.02$ TeV with nCTEQ15 nPDFs (red)
 are shown for a leading jet $p_T$ cut of 20 GeV \cite{Citron:2018lsq}.}
\end{figure}

Fig.\ 2 extends these studies to the HL LHC with centre-of-mass energy
per nucleon of 5.5 TeV \cite{Citron:2018lsq}.
The reach in $x_A$ would be extended by an order of
magnitude and the reduction of the uncertainty could reach a factor of four,
if the jet $p_T$ is not only measured above 20 GeV (left), but down to 8 GeV
(right). In the complementary kinematic region of large $x_A$, the small-$x$
region of the photon PDFs, on which little is known, could be probed. This is
demonstrated with two different parametrisations, which are still both
consistent with current data. At a HE-LHC, the centre-of-mass energies per
nucleon of 10.6 TeV in PbPb and 17 TeV in pPb collisions would obviously
allow to extend the kinematic reach even further, and open heavy-quark
production would shed further light on the heavy quark content of protons
and nuclei.

\subsection{Exclusive photoproduction}

When the hadrons on both sides are tagged or separated from the central hard
event by a rapidity gap, photon-photon collisions lead to the exclusive
production of lepton pairs. Their theoretical description within QED requires
not only accounting for form factors, but also absorptive effects from the
additional scattering of pomerons. Muon pairs with invariant mass above 10
GeV have been measured by ATLAS not only in ultraperipheral, but also
peripheral and central PbPb collisions \cite{Aaboud:2018eph}. For the former,
the leptons are mostly back-to-back as expected, and the acoplanarity
distribution ($\alpha$, top) and lepton energy imbalance ($A$, bottom) agree
well with the STARlight predictions in Fig.\ 3. In central collisions, however,
\begin{figure}
 \includegraphics[width=\textwidth]{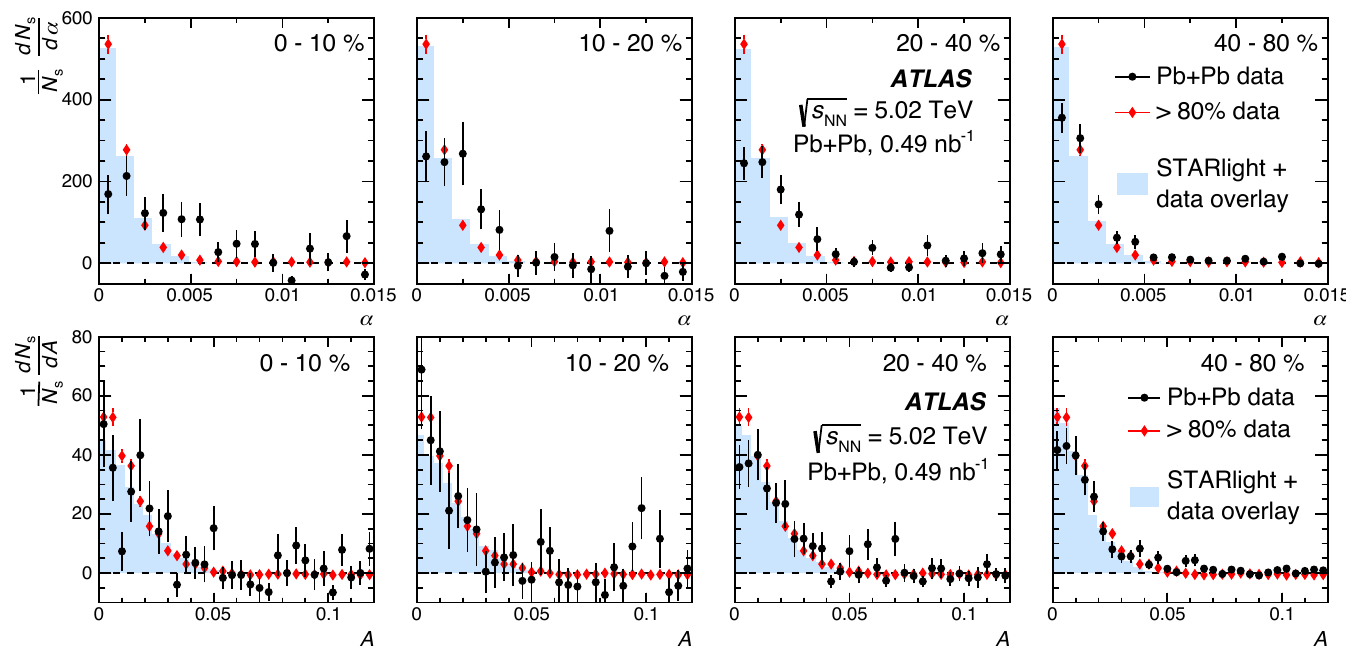}
 \caption{Acoplanarity ($\alpha$, top) and lepton energy imbalance ($A$,
 bottom) as a function of centrality for dimuon pairs with pair mass above
 10 GeV observed in the ATLAS detector \cite{Aaboud:2018eph}.}
\end{figure}
the acoplanarity peak at $\alpha=0$ is reduced, indicating electromagnetic
rescattering in the created QGP, while $A$ remains unchanged, so that no
significant energy loss through bremsstrahlung occurs. A HL LHC would reach
higher invariant masses of up to 100 GeV, corresponding to photon-photon
interactions near or even in the nuclei and thus increased interactions with
the QGP and/or nuclear magnetic field \cite{Citron:2018lsq}. At low mass,
electron pairs are expected to interact more than muons or even taus. At
higher order, also the production of four leptons can be considered
\cite{fk}.

Exclusive dijets are produced not only by photon, but also pomeron
interactions and could therefore in the future be used to determine
for the first time diffractive PDFs of nuclei \cite{Guzey:2016tek}.
At leading twist, diffraction can be related to nuclear shadowing,
and more evidence for the latter has recently
been obtained by ALICE from exclusive production of $\rho$, $J/\psi$ and
$\psi'$ mesons \cite{mb}. Their different masses would allow to probe in
the future more precisely the evolution of nuclear shadowing with $Q^2$
\cite{Guzey:2013xba}, as is shown in Fig.\ 4. It assumes that a total
integrated luminosity of 13 nb$^{-1}$ could be reached with yearly PbPb runs
at the end of 2021-2023 and 2027-2029. These measurements would be
particularly interesting
\begin{figure}
 \centering
 \includegraphics[width=0.65\textwidth]{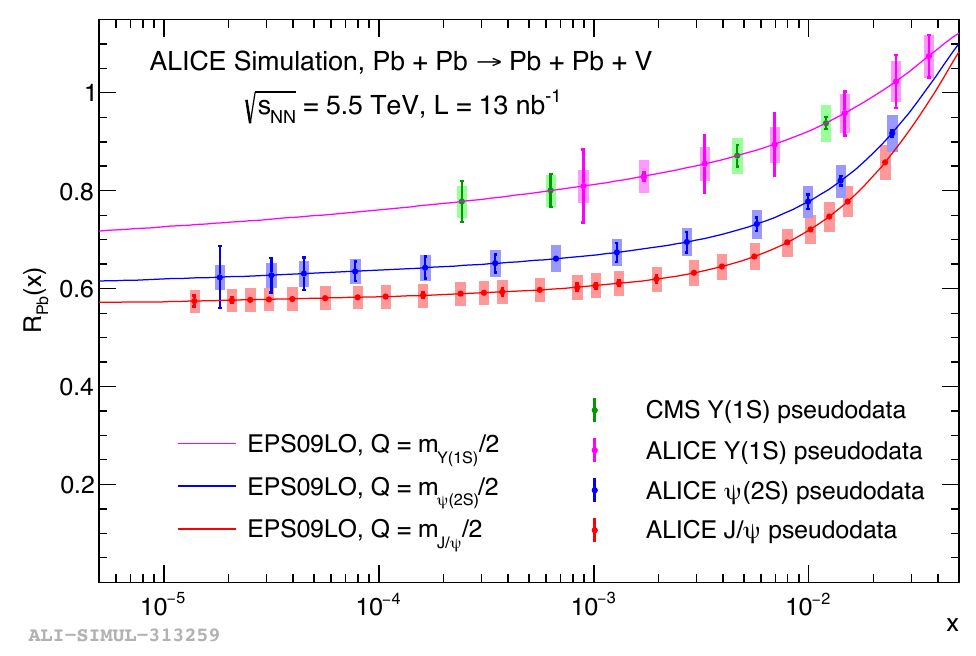}
 \caption{Pseudodata projections for the nuclear suppression factor by ALICE
 and CMS measured with the photoproduction of three heavy vector mesons in
 PbPb ultraperipheral collisions. The pseudodata points are derived from
 EPS09-based photoproduction cross section projections \cite{Guzey:2013xba}.}
\end{figure}
in view of establishing deviations from DGLAP and evidence for BFKL evolution
or saturation effects.
Finally, the production of $W$-boson and top-quark pairs as well as Higgs
bosons would allow to search more precisely for deviations of the electroweak
couplings of these particles from the SM predictions \cite{de}.

\section{SM physics with photons at the EIC}

Plans to build an EIC at either BNL or JLAB are well advanced. It would allow
for a diverse physics program impacting nuclear, heavy-ion and high-energy
physics with studies of sea quark and gluon distributions, their spins and the
emergence of nuclear properties through electromagnetic, i.e.\ photon,
interactions \cite{ea}.

As two examples, we discuss here the impact of inclusive jets and dijets in
deep-inelastic scattering (DIS) and photoproduction on the determination of
nuclear PDFs. Both processes have recently been calculated at approximate
NNLO (aNNLO). Fig.\ 5 (top left) shows the $p_T$ distribution of inclusive
\begin{figure}[t!]
 \includegraphics[width=0.49\textwidth]{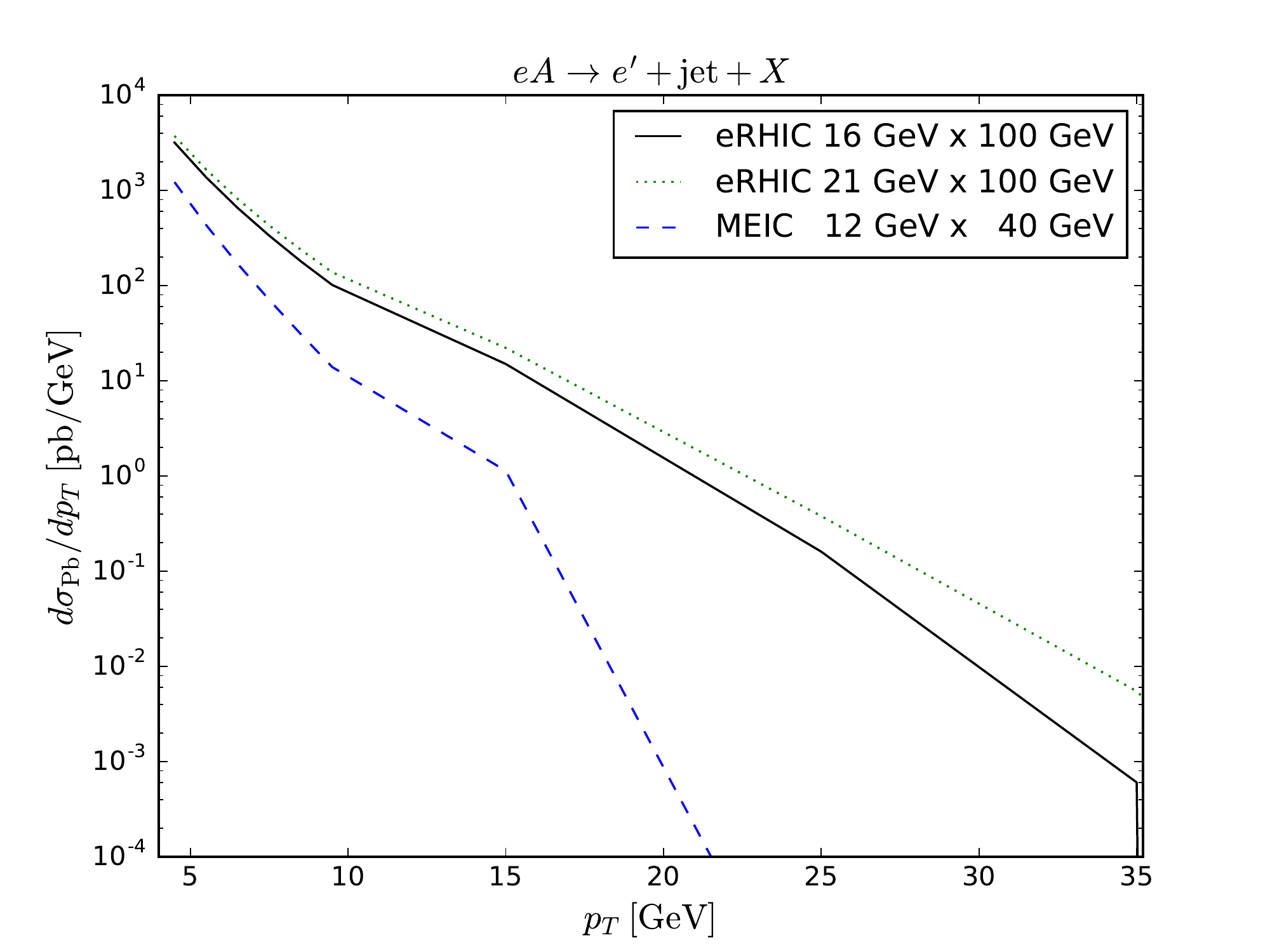}
 \includegraphics[width=0.49\textwidth]{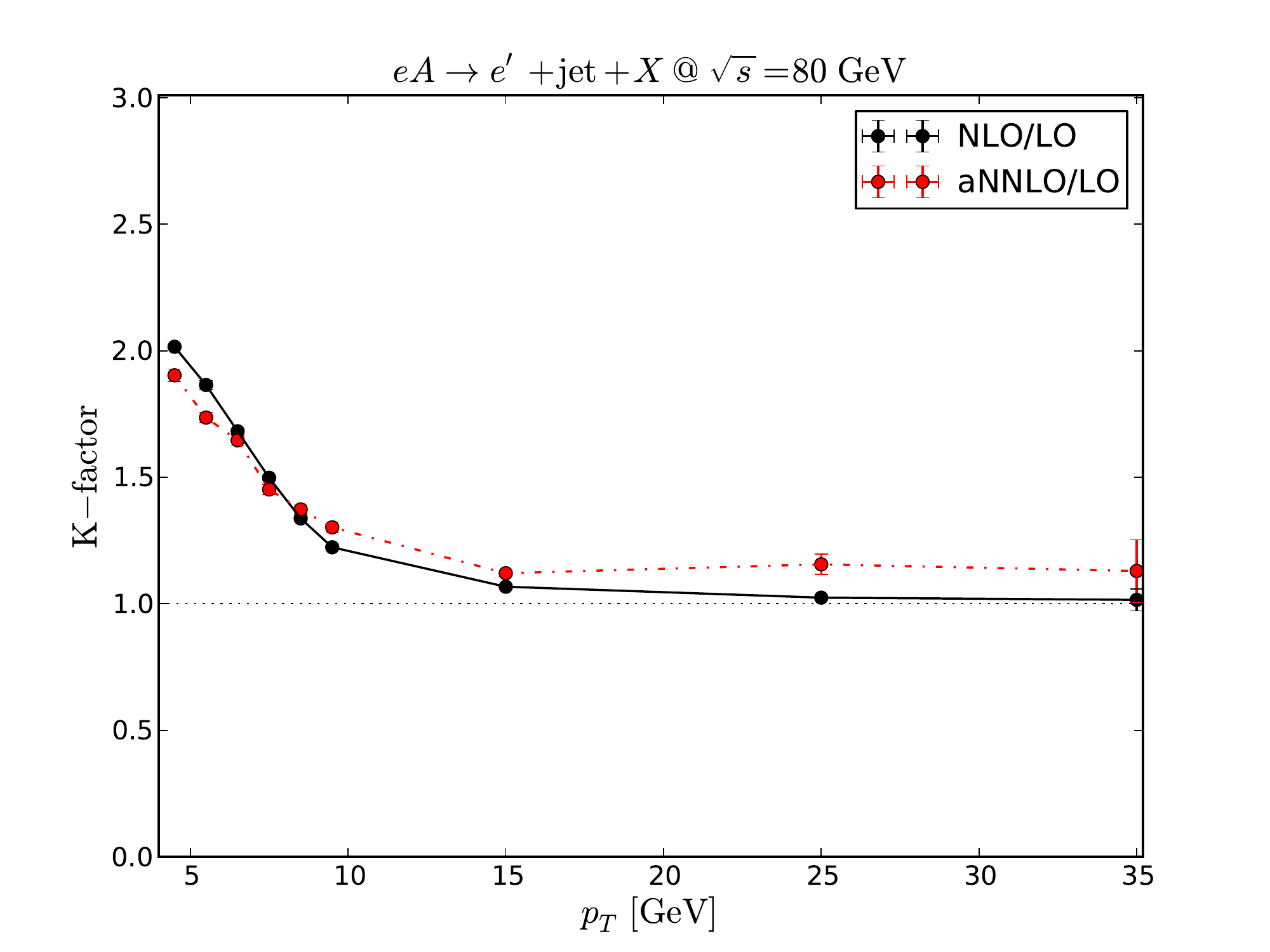}
 \includegraphics[width=0.49\textwidth]{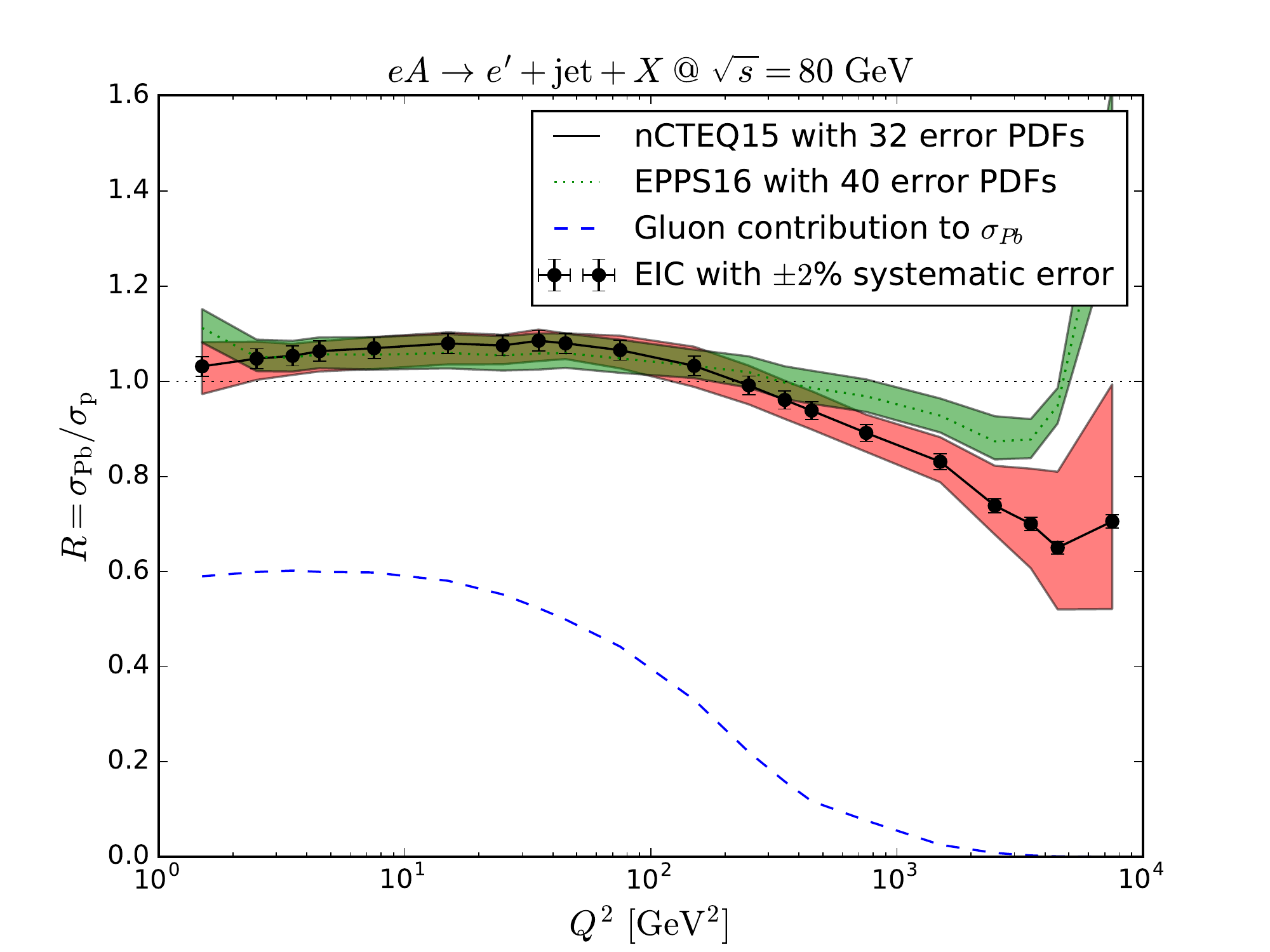}\,\,\,
 \includegraphics[width=0.49\textwidth]{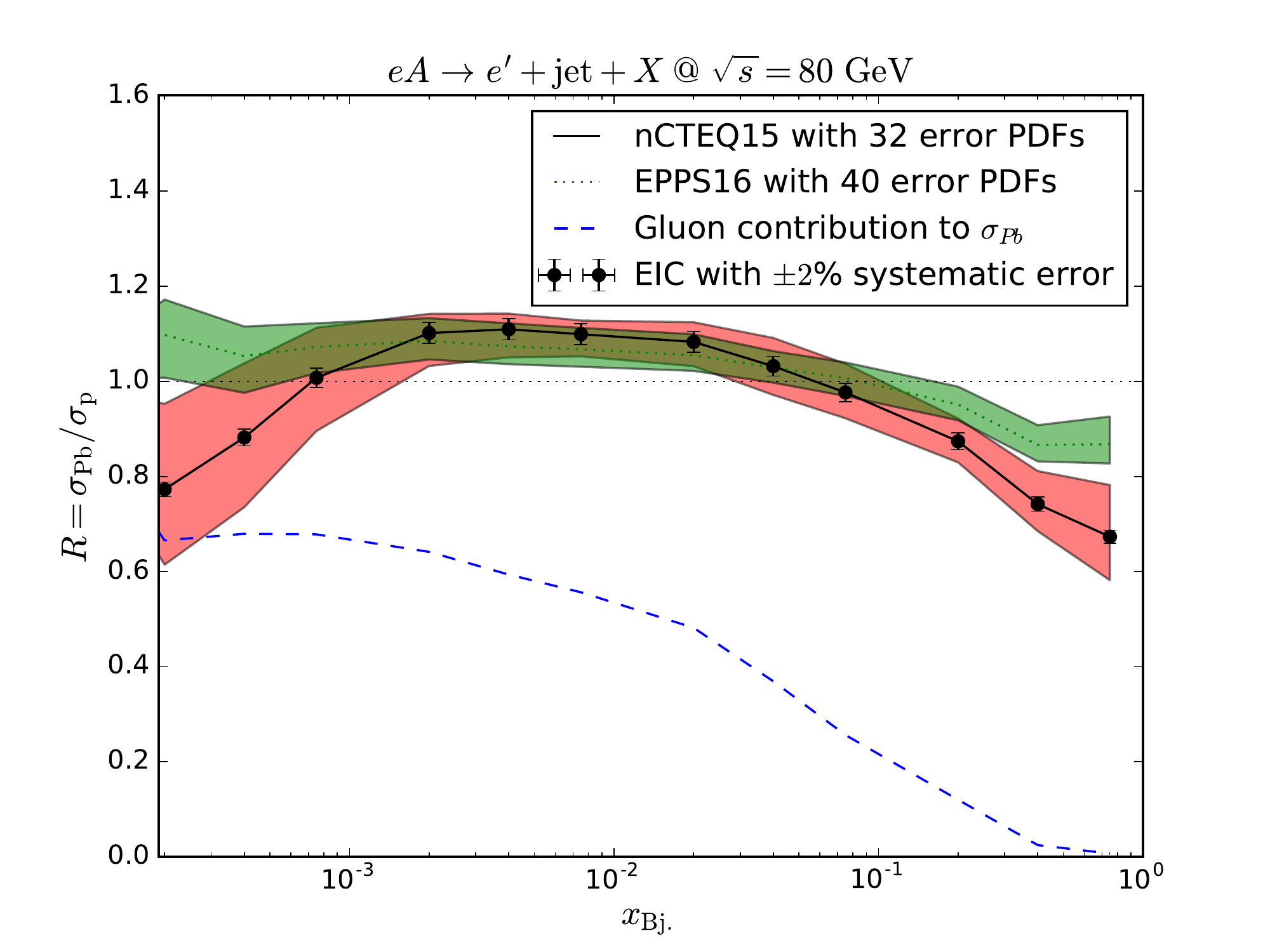}
 \caption{Top left: $p_T$ distribution of inclusive jets in DIS for different
 EIC designs. Top right: $K$ factors as a function of $p_T$ at NLO and aNNLO.
 Bottom left: $Q^2$ evolution as predicted by nCTEQ15 and EPPS16. Bottom right:
 Nuclear PDF uncertainty bands as a function of the reconstructed parton
 momentum fraction in the lead nucleus \cite{Klasen:2017kwb}.}
\end{figure}
jets in DIS for different EIC designs, where the eRHIC option with a 21 GeV
electron and a 100 GeV per nucleon ion beam allows to reach $p_T$ values of up
to 35 GeV. The $K$ factors
as a function of $p_T$ at NLO and aNNLO (top right) are very similar, which
demonstrates good perturbative stability, as are the $Q^2$ evolutions
predicted by nCTEQ15 and EPPS16 (bottom left), both based on DGLAP. However,
the two nuclear PDF uncertainty bands do not overlap at $x$ below $10^{-3}$,
demonstrating the potential EIC impact \cite{Klasen:2017kwb}.

Similar distributions are shown in Fig.\ 6 for dijet photoproduction.
\begin{figure}[t!]
 \includegraphics[width=0.49\textwidth]{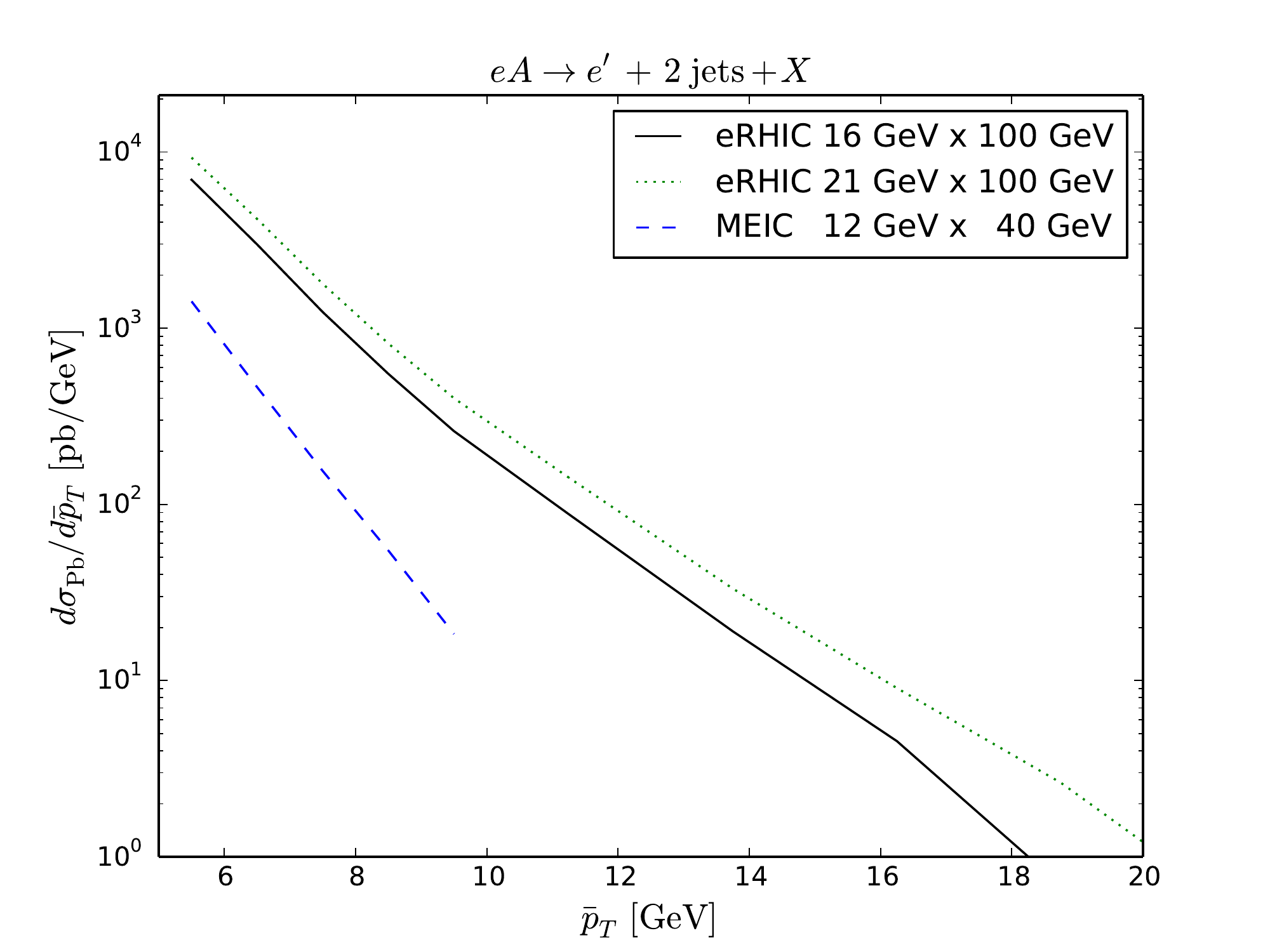}
 \includegraphics[width=0.49\textwidth]{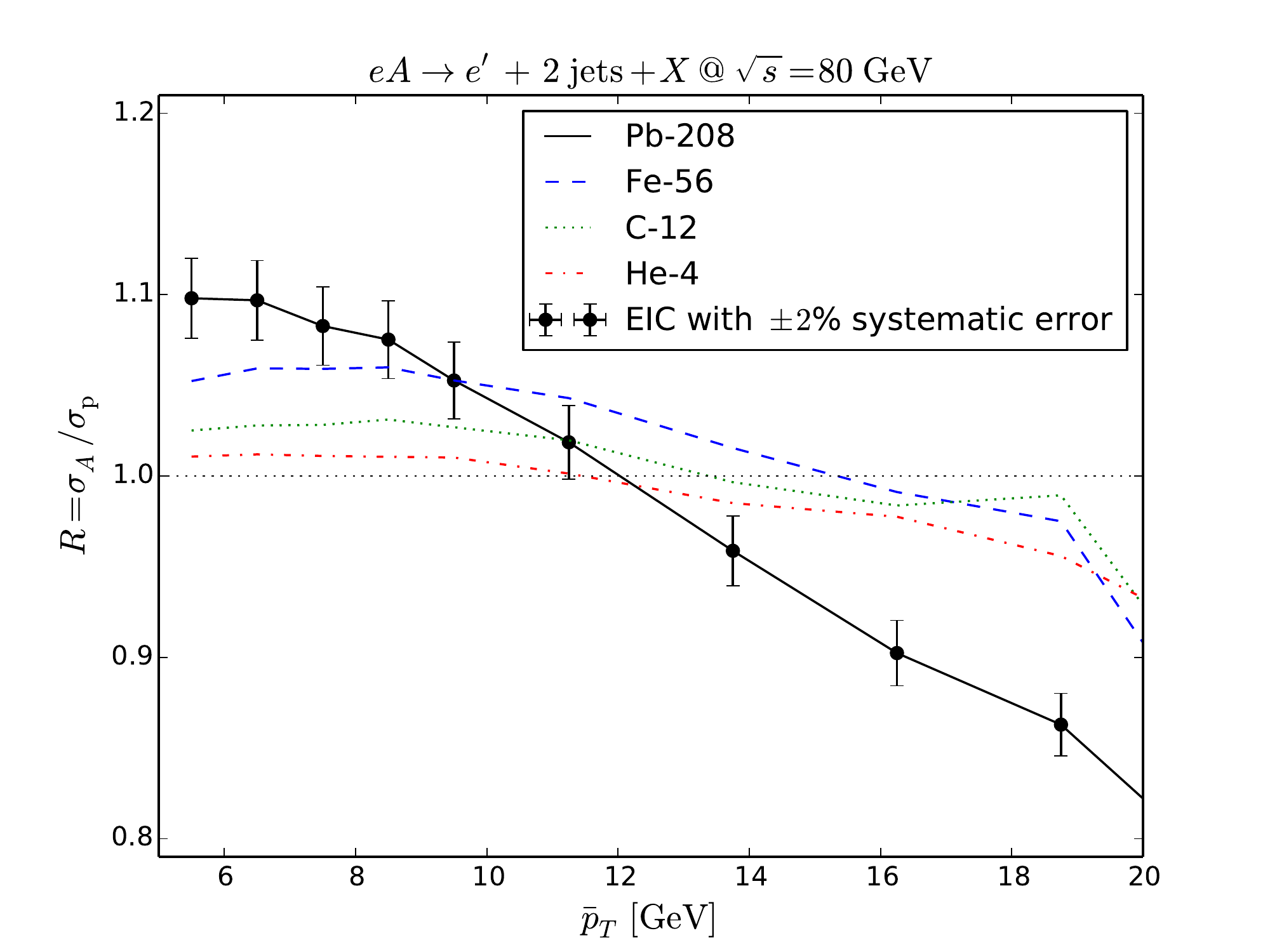}
 \includegraphics[width=0.49\textwidth]{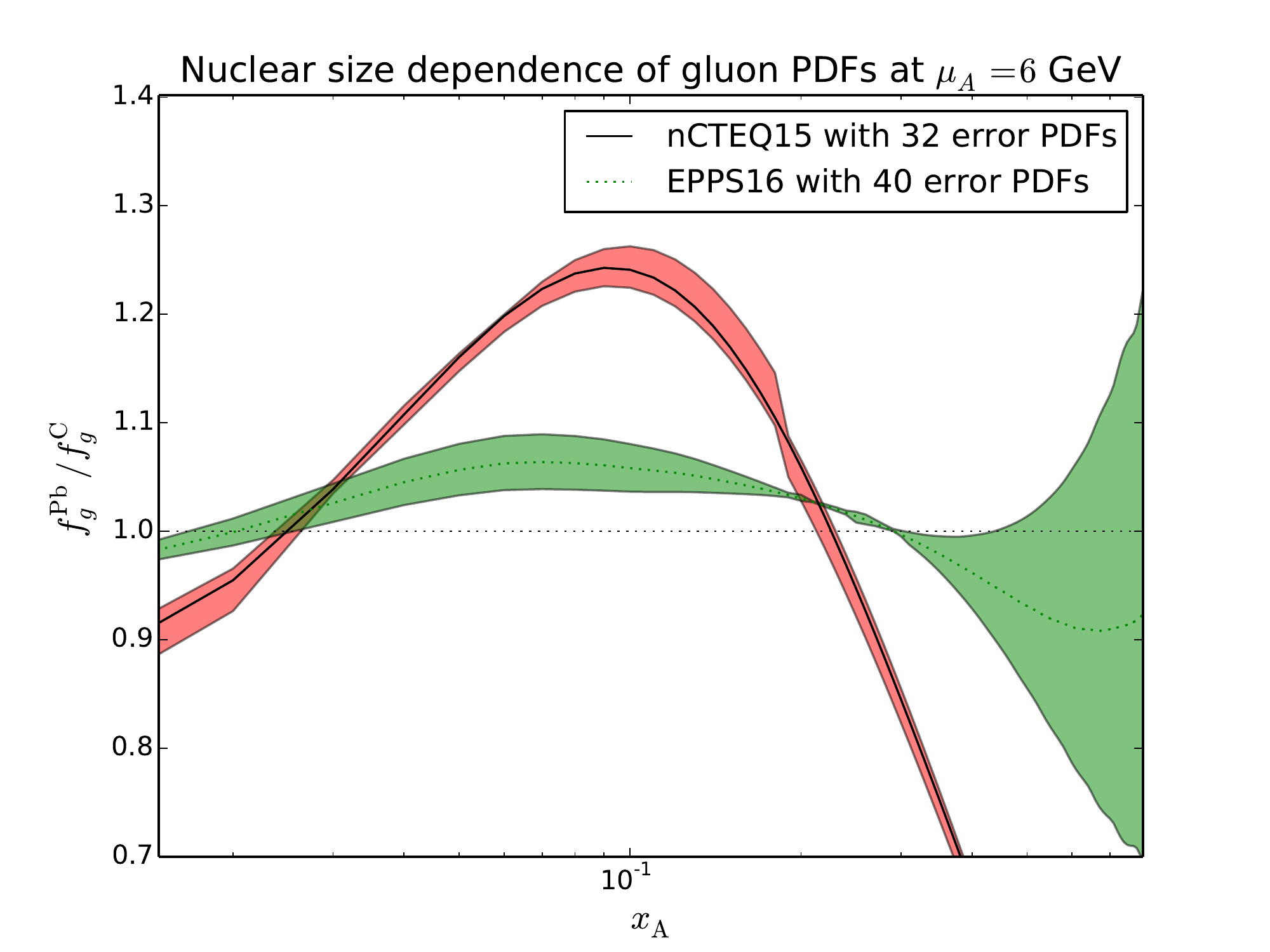}\,\,\,
 \includegraphics[width=0.49\textwidth]{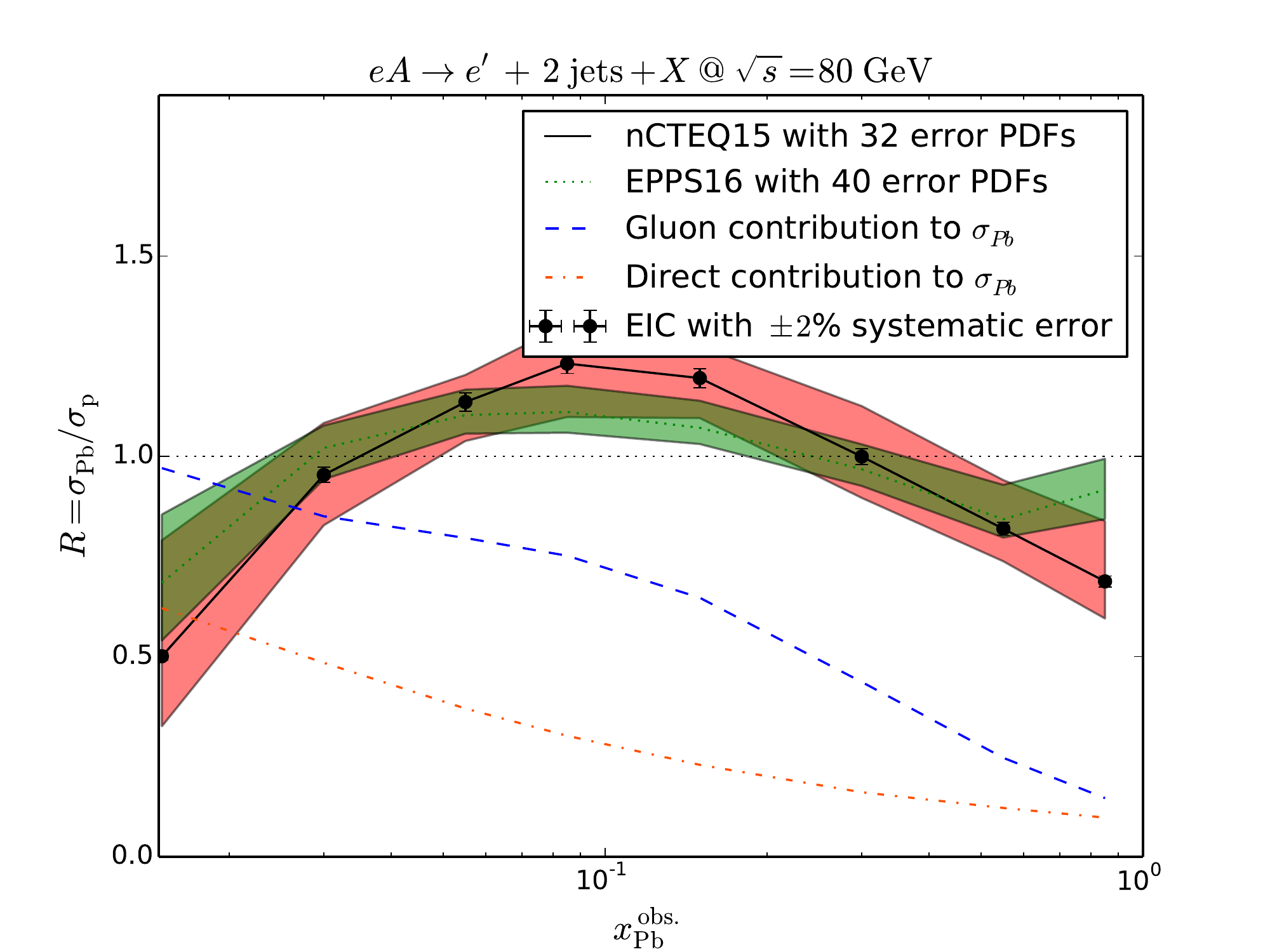}
 \caption{Top left: Average $p_T$ distribution of dijets in photoproduction for
 different EIC designs. Top right: Ratios of different nuclear over free
 proton cross sections as a function of $p_T$. Bottom left: Ratios of lead and
 carbon over free proton cross sections as a function of the reconstructed
 parton momentum fraction in the nucleus. Bottom right: Nuclear PDF
 uncertainty bands as a function of the reconstructed parton momentum fraction
 in the lead nucleus \cite{Klasen:2018gtb}.}
\end{figure}
The average $p_T$ of the two jets is now restricted to below 20 GeV (top left).
The nuclear modifications depend strongly on the nucleus (top right) and are
modelled differently by nCTEQ15 and EPPS16 (bottom left). Due to the reduced
partonic centre-of-mass energy, the nuclear PDF sensitivity does not extend
to $x$ values below $10^{-2}$.
An alternative process to constrain in particular the nuclear gluon density
would again be exclusive quarkonium photoproduction, also at the EIC \cite{al}.

\section{BSM physics with photons at the high-luminosity and high-energy LHC}

The searches for anomalous couplings of weak gauge or Higgs bosons and top
quarks already briefly mentioned above are mainly motivated by the hierarchy
and unification problems of the SM. Another major motivation for BSM searches
comes from dark matter (DM), whose existence is largely undisputed, but whose
nature remains to be elucidated. We therefore focus in this section on three
different DM candidates, all related to photons: weakly interacting massive
particles (WIMPs) and their future constraints from monophotons;
prospects for dark photon searches; and axion-like particle (ALP)
contributions to light-by-light scattering at the HL and HE LHC.

\subsection{Future dark matter searches with monophotons}

Monophoton searches at the LHC can be competitive to other processes, in
particular monojets, when DM is part of an electroweak multiplet, since photons
induce a different dependence on model parameters like the electroweak
representation or mass splitting. For example, DM is part of a Higgsino
triplet ($\chi^0,\chi^\pm$) in anomaly-meadiated SUSY-breaking models
\cite{Fuks:2011dg} and of scalar or fermion singlets, doublets or triplets in
minimal DM models with a SM mediator \cite{Klasen:2013btp}.
Even when DM and its charged multiplet partners have identical masses at
tree-level, electroweak loops always induce a mass splitting, e.g.\ of
$m_{\chi^\pm}=m_{\chi^0}+165$ MeV for triplets, making the neutral partner
lighter and the heavier ones decay like $\chi^\pm\to\chi^0$ + soft charged
pions. The DM particle itself is usually stabilised against decay into SM
particles by assuming a symmetry like $R$ parity or $U(1)_{B-L}$. The observed
thermal relic density can then be obtained for masses $m_{\chi^0}\leq3$ TeV.

DM signals from monophoton searches at the LHC not only have to be
discriminated against the irreducible background $Z(\to\nu\nu)\gamma$, but also
from $W/Z$ + jet, $tt$, $ZZ/WW$ production with electrons or jets faking
photons. This is achieved with kinematic cuts like $\not{\!E}_T>150$ GeV,
$p_T^\gamma>150$ GeV, $|\eta^\gamma|<2.37$ and a photon isolation from
$\not{\!E}_T$ by $\Delta\phi>0.4$. The LHC can then set stronger limits
than LEP ($m_{\chi^0}>90$ GeV) as shown in Fig.\ 7 (left), reaching DM
\begin{figure}[t!]
 \centering
 \includegraphics[width=0.505\textwidth]{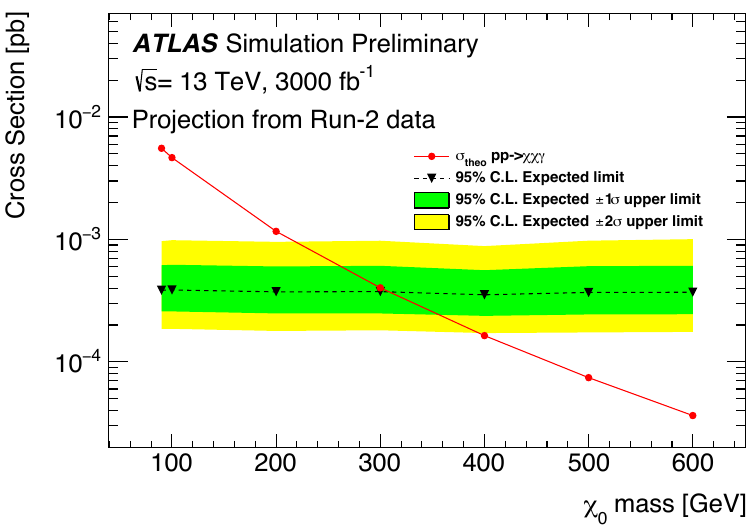}
 \includegraphics[width=0.48\textwidth]{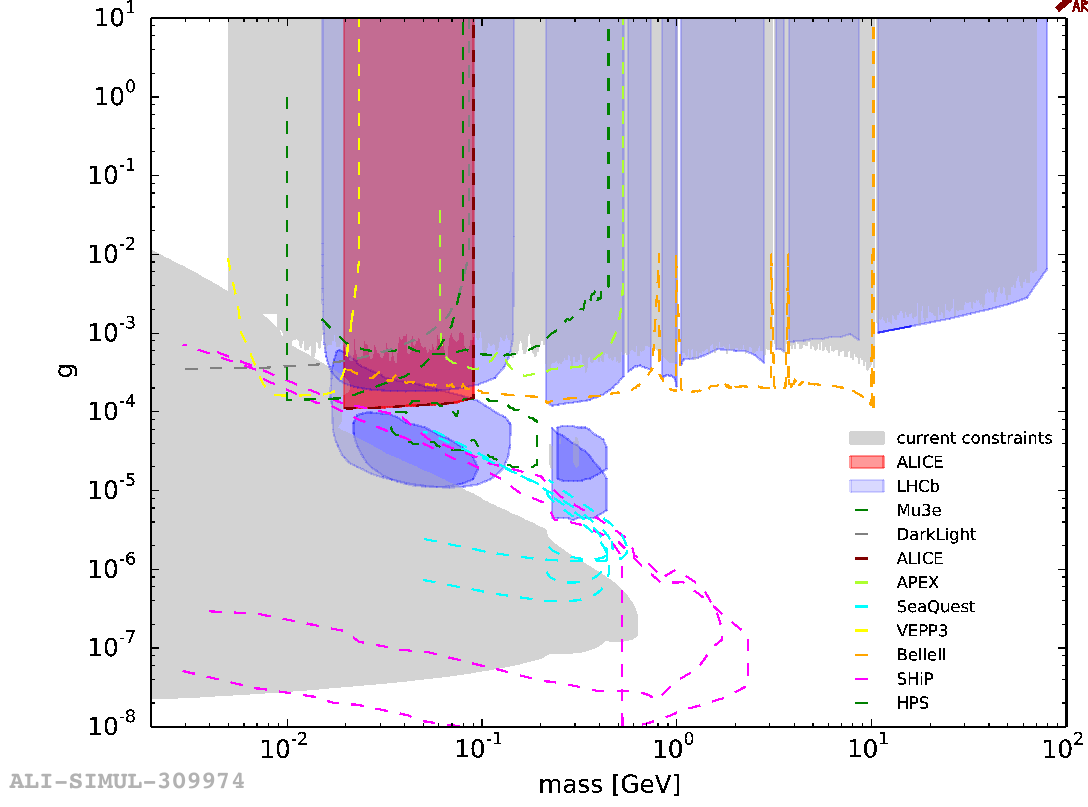}
 \caption{Left: Expected upper limits at 95\% C.L.\ on the production cross
 section of dark matter as a function of $\chi^0$ mass in monophoton final
 state. Results are shown for an integrated luminosity of 3 ab$^{-1}$. The
 red line shows the theoretical cross section \cite{CidVidal:2018eel}.
 Right: 90\% of C.L.\ constrained by ALICE and LHCb in HL LHC era. Constraints
 by ALICE are based on 6 pb$^{-1}$, 0.3 pb$^{-1}$, 10 nb$^{-1}$, 0.3 pb$^{-1}$,
 and 3 nb$^{-1}$ of pp, pPb, and PbPb collisions at 0.5 T, and pPb and PbPb
 collisions at 0.2 T and by LHCb are based on 15 fb$^{-1}$
 \cite{Citron:2018lsq}.}
\end{figure}
masses of 300 GeV for an integrated luminosity of 3 ab$^{-1}$
\cite{CidVidal:2018eel}.

\subsection{Prospects for dark photon searches}

Dark photons $A'$ from $U(1)$ gauge extensions of the SM have gained in
popularity
as the neutral SM gauge and Higgs bosons have become more and more excluded as
mediators for WIMP DM in the mass region between a few GeV and TeV. They are
parametrised by their mass, obtained from spontaneous symmetry breaking, and
mixing parameter $g$ with the SM photon. ALICE has searched for possible
decays of $\pi^0\to \gamma A'(\to e^+e^-)$ by examining the electron-positron
invariant mass between 20 and 90 MeV in pp and pPb collisions, and its upgrade
will greatly improve the efficiency. LHCb has good capabilities to measure
muon pairs and thus searches for prompt-like and long-lived dark photons
produced in pp collisions and decaying as
$A'\to\mu^+\mu^-$ between 214 MeV and 70
GeV. As Fig.\ 7 (right) shows, smaller couplings $g$ will be probed at the
HL LHC, closing potentially the wedge in the 20 to 90 MeV mass region.

\subsection{BSM perspectives in light-by-light scattering}

The $Z^4$ enhancement in PbPb collisions at the LHC leads to $4.5\cdot10^7$
more initial photon pairs than obtained in pp collisions, albeit with a softer
spectrum. Both ATLAS and CMS have now observed light-by-light scattering, i.e.\
the exclusive production of diphotons in UPCs. Apart from model-independent
searches for anomalous couplings, they allow in particular to hunt for light
ALPs, which arise in solutions of the strong CP problem, through the
identification of invariant mass peaks that should be clearly visible above
the steeply falling QED background, as shown in Fig.\ 8 (left).
\begin{figure}[t!]
 \includegraphics[width=0.485\textwidth]{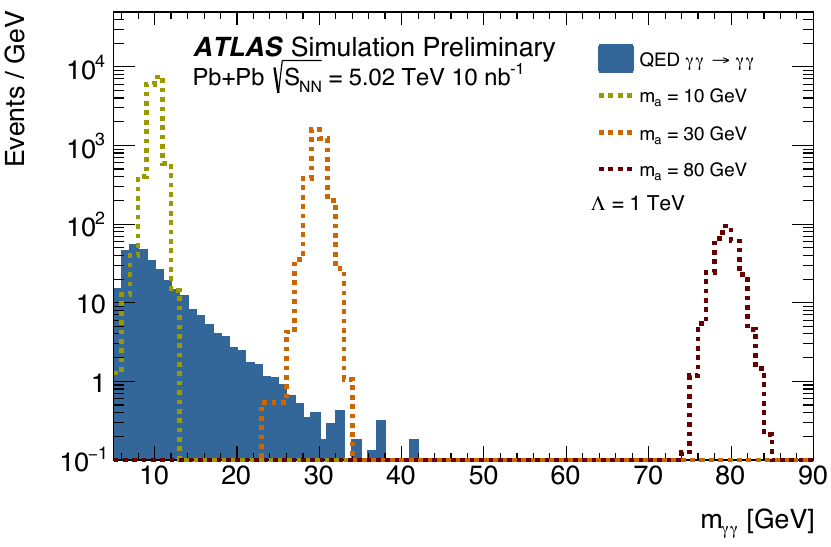}
 \includegraphics[width=0.455\textwidth]{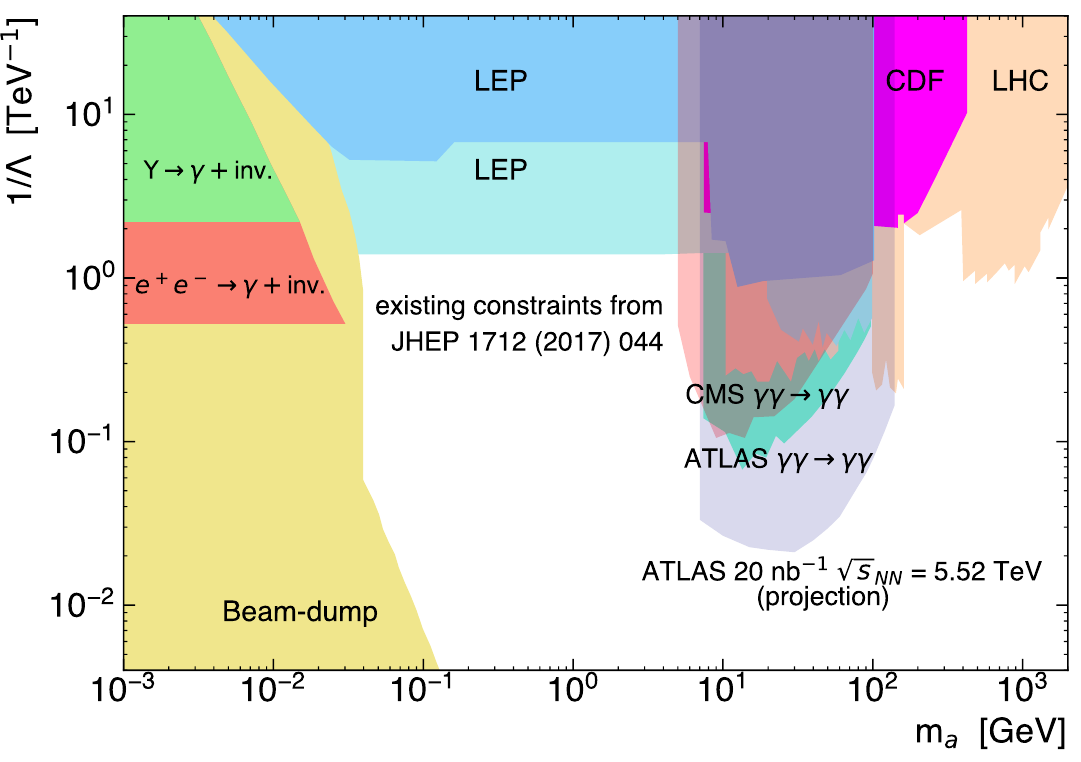}
 \caption{Left: Mass distribution for the ALP signal shown for three values of
 the ALP mass $m_a = 10$, 30 and 80 GeV (in red). Also shown (in blue) is the
 QED background. All ALP mass points are generated with $\Lambda =
 1$ TeV. Right: Compilation of exclusion limits obtained by different
 experiments. In light grey, the projected ATLAS 20 nb$^{-1}$ limit at
 $\sqrt{s_{NN}} = 5.52$ TeV is presented \cite{Citron:2018lsq}.}
\end{figure}
Upper limits can then be set on the product of the production cross section
and decay branching ratio into diphotons. In Fig.\ 8 (right), existing
exclusion limits on the ALP coupling, $1/\Lambda$, as a function of its mass
$m_a$ are supplemented with a projected ATLAS limit derived from PbPb
collisions at 5.52 TeV. These results demonstrate that heavy-ion collisions
have unique sensitivity to ALP searches in the mass range from 7 to 140 GeV
\cite{Citron:2018lsq}.

Many theory papers have been written on BSM searches at photon colliders
around the year 2000 in view of the expected construction of a linear collider.
As an example, the testable scale of non-commutative QED was foreseen at
$\Lambda_{\rm NC}\geq1.5\,\sqrt{s_{ee}}$.
However, a few studies have also been performed for the LHC. E.g., monopole
mass limits of $M<n\cdot7.4,\,10.5,\,19$ TeV were expected for $J_M=0,\,1/2,\,
1$ at $\sqrt{s_{pp}}=7$ TeV, and limits of $M_{\rm Pl.}\geq5...8\,
\sqrt{s_{\gamma\gamma}}$ were predicted for $D=4+(2,4,6)$ dimensional gravity.
With the discovery of the Higgs boson, ``unparticles''
are now all but forgotten \cite{Citron:2018lsq}. Nevertheless, contrary to
standard SUSY LHC searches, photon-photon collisions might indeed be sensitive
in compressed mass scenarios where e.g.\ $m_{\tilde{l}}\sim m_{\tilde{\chi}^0}$
\cite{lhl}. The search for monopoles with ATLAS, where the current mass limit
from 13 TeV pp collisions lies at 2 TeV, has proven more difficult than
expected, but is ongoing with the dedicated experiment MoEDAL and might
in the future benefit from the enhanced photon luminosity in PbPb collisions
\cite{og}.

\section{Conclusion}

In conclusion, we have tried to present a balanced and realistic discussion of
physics opportunities with photons at future colliders, focusing on either
existing (SM) physics at colliders with advanced funding decisions or on BSM
physics at the HL LHC already under construction. Particular attention has
been spent on the unique potential of photons to constrain the proton and in
particular nuclear structure at high energy as well as their role in searches
for DM, currently our clearest hint of physics beyond the SM. Photons also
play of course a crucial role in astroparticle physics, but a thorough
discussion of cosmic rays, the upcoming CTA telescope and the fascinating
perspectives of multimessenger astronomy were unfortunately beyond the scope
of this conference summary talk.

\section*{Acknowledgements}

The author thanks the organisers of the Photon 2019 conference for the
kind invitation, V.\ Guzey, T.\ Jezo, C.\ Klein-B\"osing, F.\ K\"onig,
K.\ Kovarik, H.\ Poppenborg, J.\ Potthoff and J.P.\ Wessels for their
collaboration, and the DFG for financial support under contract KL 1266/9-1.


\end{document}